\documentclass [12pt]{article}

\usepackage{amsmath,amssymb,amsthm,epsfig}
\usepackage{color}

\def\ulambda{{\underline{\lambda}}}
\def\Xom{{\mathcal L}^2_{\ulambda}(\Omega)}
\def\Yom{{\mathcal H}^{1/2}_\ulambda(\Omega)}
\def\Zom{{\mathcal H}^{1}_\ulambda(\Omega)}
\def\dYom{\dot{\mathcal H}^{1/2}_\ulambda(\Omega)}
\def\dZom{\dot{\mathcal H}^{1}_\ulambda(\Omega)}

\def\N{{\mathbb N}}

\def\eqnn{\begin{eqnarray*}}
\def\eeqnn{\end{eqnarray*}}
\def\eqn{\begin{eqnarray}}
\def\eeqn{\end{eqnarray}}

\def\prf{\begin{proof}}
\def\endprf{\end{proof}}

\theoremstyle{plain}
\newtheorem{theorem}{Theorem}%[section]

\newtheorem{lemma}[theorem]{Lemma}

\numberwithin{equation}{section}

\begin{document}

\title{On the well-posedness of the semi-relativistic
Schr\"odinger-Poisson system}

\date{}

\maketitle

\centerline{Walid Abou Salem}

\centerline{Department of Mathematics and Statistics, 
University of Saskatchewan,}

\centerline{Saskatoon, SK, S7N 5E6, Canada}
 
\centerline{e-mail: walid.abousalem@usask.ca}

\bigskip

\centerline{Thomas Chen}

\centerline{Department of Mathematics, University of
Texas at Austin,}

\centerline{Austin, TX, 78712, USA}

\centerline{e-mail: tc@math.utexas.edu}
 
\bigskip

\centerline{Vitali Vougalter}

\centerline{University of Cape Town, Department of Mathematics
and Applied Mathematics,}
 
\centerline{Private Bag, Rondebosch 7701, South Africa}

\centerline{e-mail: Vitali.Vougalter@uct.ac.za}

\medskip

\bigskip
\bigskip
\bigskip

\noindent {\bf Abstract.} 
We show global existence and uniqueness of strong solutions
for the Schr\"odinger-Poisson system
in the repulsive Coulomb case with   relativistic kinetic energy.

\bigskip
\bigskip


\bigskip


\newpage

\section{Introduction}

\bigskip
 
In this article, we study the global well-posedness of the semi-relativistic Schr\"odinger-Poisson system on a finite domain. This system is relevant to the description of many-body semi-relativistic quantum particles in the mean-field limit (for instance, in heated plasma), when the particles move with extremely high velocities. Consider semi-relativistic quantum particles confined in
domain $\Omega\subset {\mathbb R}^{3}$ which is an open, finite volume set with 
a ${C}^{2}$ boundary. The particles interact by the
electrostatic field they collectively generate. In the mean-field limit, the density matrix that describes the {\it mixed} state of the system satisfies the Hartree-von Neumann equation
\begin{equation}
\label{eq:HartreeVonNeumann}
\begin{cases}
i\partial_t \rho(t) = [H_V,\rho(t)], \ \ x\in \Omega, \ \ t\ge 0\\
-\Delta V = n(t,x), \ \ n(t,x)=\rho(t,x,x),
\ \rho(0)=\rho_0
\end{cases},
\end{equation}
satisfying Dirichlet boundary conditions, $\rho(t,x,y)=0$ if $x$ or $y\in\partial\Omega$, for $t\geq0$.
The Hamiltonian is given by 
\eqn\label{eq:HV-def-1}
	H_{V}:=T_{m}+V(t,x)
\eeqn
where the relativistic kinetic energy operator
$T_{m}:=\sqrt{-\Delta+m^{2}}-m$ is defined via the spectral calculus. Here, 
$\Delta$ denotes the Dirichlet Laplacian on $L^{2}(\Omega)$, and $m>0$ is the 
particle mass;  see \cite{AS09, A10} for a derivation of this system of equations in the {\it non-relativistic} case. 
Since $\rho(t)$ is a positive, self-adjoint trace-class operator acting on $L^2(\Omega)$, its kernel can, for every $t\in\mathbb{R}_+$, be decomposed with respect to an  orthonormal basis of $L^2(\Omega)$. 
The kernel of the initial data $\rho_0$ can be represented in the form
\begin{equation}
\label{eq:kernelinit}
\rho_0(x,y) = \sum_{k\in {\mathbb N}} \lambda_k \psi_k(x)\overline{\psi_k(y)}
\end{equation}
where $\{\psi_k\}_{k\in {\mathbb N}}$ denotes an orthonormal basis of $L^2(\Omega)$, 
with $\psi_k|_{\partial\Omega}=0$ for all $k\in\N$, and coefficients 
\eqn
	\ulambda:=\{\lambda_k\}_{k\in {\mathbb N}}\in \ell^1
	\;, \;\;
	\lambda_k\geq0
	\; , \; \;
	\sum_k\lambda_k=1.
\eeqn 
As shown below, there exists a one-parameter family of 
complete orthonormal bases of $L^2(\Omega)$, $\{\psi_k(t)\}_{k\in {\mathbb N}}$, with $\psi_k(t)|_{\partial\Omega}=0$ for all $k\in\N$, and for $t\in\mathbb{R}_+$,  such that  the kernel of the solution $\rho(t)$ to \eqref{eq:HartreeVonNeumann} can be represented as 
\begin{equation}
\label{eq:kernel}
\rho(t,x,y) = \sum_{k\in {\mathbb N}} \lambda_k \psi_k(t,x)\overline{\psi_k(t,y)}.
\end{equation}
Notably, the coefficients $\ulambda$
are {\em independent} of $t$, and thus the same as those in $\rho_0$.
Substituting (\ref{eq:kernel}) in (\ref{eq:HartreeVonNeumann}), the one-parameter family of orthonormal vectors  $\{\psi_k(t)\}_{k\in {\mathbb N}}$ is seen to satisfy the semi-relativistic Schr\"odinger-Poisson system
\begin{equation}
\label{sch}
i\frac{\partial \psi_{k}}{\partial t}=T_{m}\psi_{k}+V\psi_{k}, \quad k\in 
{\mathbb N}
\end{equation}  
\begin{equation}
\label{p}
-\Delta V[\Psi]=n[\Psi], \quad \Psi:=\{\psi_{k}\}_{k=1}^{\infty},
\end{equation} 
\begin{equation} 
\label{n}
n[\Psi(x,t)]=\sum_{k=1}^{\infty}\lambda_{k}|\psi_{k}|^{2},
\end{equation} 
with initial data $\{\psi_k(0)\}_{k=1}^\infty$.
The potential function $V[\Psi]$ solves the Poisson equation
(\ref{p}). On both $V[\Psi]$ and $\psi_k(t)$, for all $k\in\N$, we impose Dirichlet boundary conditions
\begin{equation}
\label{dV}
\psi_k(t,x) \; , \; \;
V(x,t)=0, \ t\geq 0, \ \forall x\in \partial \Omega.
\end{equation}
As we show in Lemma \ref{lm:6}, below, 
solutions of (\ref{sch})-(\ref{n})  preserve the orthonormality
of  $\{\psi_k(t)\}_{k\in {\mathbb N}}$.

The state space for the Schr\"odinger-Poisson system  is given by
$$
{\cal L}:=\{ (\Psi, \ulambda) \ | \ \Psi=\{\psi_{k}\}_{k=1}^{\infty}\subset
H_{0}^{\frac{1}{2}}(\Omega)\cap H^{1}(\Omega) \; \; is \ a \ complete \ 
orthonormal \ system \ in \ L^{2}(\Omega), 
$$
$$
\ulambda=\{\lambda_{k}\}_{k=1}^{\infty}\in \ell^1, \quad \lambda_{k}\geq 0, \ k\in
{\mathbb N}, \quad \sum_{k=1}^{\infty}\lambda_{k}\int_{\Omega}|\nabla \psi_{k}|^{2}
dx<\infty \}.
$$
For fixed $\ulambda\in \ell^1, \ \ \lambda_k>0,$ and for sequences of square integrable functions 
$\Phi:=\{\phi_{k}\}_{k=1}^{\infty}$ and 
$\Psi:=\{\psi_{k}\}_{k=1}^{\infty},$ we define the inner product
$$
(\Phi,\Psi)_{\Xom}:=\sum_{k=1}^{\infty}\lambda_{k}(\phi_{k}, 
\psi_{k})_{L^{2}(\Omega)},
$$ 
which induces the norm 
$$
\|\Phi\|_{\Xom}:=(\sum_{k=1}^{\infty}\lambda_{k}{\|\phi_{k}\|
_{L^{2}(\Omega)}^{2}})^{\frac{1}{2}},
$$
and we introduce the corresponding Hilbert space
$$
\Xom:=\{\Phi=\{\phi_{k}\}_{k=1}^{\infty} \ | \ \phi_{k}\in
L^{2}(\Omega), \ \forall \ k\in {\mathbb N}, \ \|\Phi\|_{\Xom}<\infty \}.
$$
%\bigskip
%
Our main result is as follows.

\bigskip

\begin{theorem}\label{thm:1}
For every initial state $(\Psi(x,0),\ulambda)\in {\cal L},$
there is a unique mild solution $\Psi(x,t)$, $t\in [0,\infty)$, of 
(\ref{sch})-(\ref{n}) with $(\Psi(x,t),\ulambda)\in {\cal L}$, 
which is also a unique strong global solution in $\Xom$.
\end{theorem}

\bigskip

Establishing the global well-posedness of the Schr\"odinger-Poisson system 
plays a crucial role in proving the existence and nonlinear stability of 
stationary states, i.e. the nonlinear bound states of the Schr\"odinger-Poisson system, 
which was done in the nonrelativistic case in \cite{BM91,MRW02}. The problem in one dimension was treated in \cite{S91}. The 
semiclassical limit of the Schr\"odinger-Poisson system with the relativistic 
kinetic energy was studied in the recent article \cite{AMS08}. Global 
well-posedness for a single semi-relativistic Hartree equation in 
${\mathbb R}^{3}$ was established in \cite{L07}. In the present work, we deal
with the infinite system of equations in a finite volume set with Dirichlet
boundary conditions, and, as distinct from \cite{L07}, we do not use the 
regularization of the Poisson equation. Moreover, both the results of \cite{L07} and 
Theorem \ref{thm:1} above do not rely on Strichartz type estimates.

\bigskip

 %%%%%%%%%%%%%%%%%%%%%%%%%%%%%%%%%%%%%%%%%%%%%%%%%%%%

\setcounter{equation}{0}

\section{Proof of global well-posedness}

\bigskip

We make a fixed choice of $\ulambda=\{{\lambda}_{k}\}_{k=1}^{\infty}\in\ell^1$, with
$\lambda_k>0$ and  $\sum\lambda_k=1$,  denoting the sequence of coefficients determined by the initial data $\rho_0$ of the Hartree-von Neumann equation \eqref{eq:HartreeVonNeumann} via \eqref{eq:kernel}, for $t=0$. We note that we require all $\lambda_k>0$ to be positive for the subsequent analysis. This does not lead to any loss of generality since by density arguments, any $\rho_0$ (and likewise $\rho(t)$) can be approximated arbitrarily well by an expansion 
of the form \eqref{eq:kernelinit}, respectively \eqref{eq:kernel}, with  $\lambda_k>0$.

We introduce inner products $(\cdot,\cdot)_{\Yom}$ and
$(\cdot,\cdot)_{\Zom}$ which induce the generalized inhomogenous Sobolev norms
$$
\|\Phi\|_{\Yom}:=(\sum_{k=1}^{\infty}\lambda_{k}{\|\phi_{k}\|
_{H^{\frac{1}{2}}(\Omega)}^{2}})^{\frac{1}{2}} \; \; {\rm and} \; \; 
\|\Phi\|_{\Zom}:=(\sum_{k=1}^{\infty}\lambda_{k}{\|\phi_{k}\|
_{H^{1}(\Omega)}^{2}})^{\frac{1}{2}},
$$
and define the corresponding Hilbert spaces 
$$
\Yom:=\{\Phi=\{\phi_{k}\}_{k=1}^{\infty} \ | \ \phi_{k}\in
H_{0}^{\frac{1}{2}}(\Omega), \ \forall \ k\in {\mathbb N}, \ 
\|\Phi\|_{\Yom}<\infty \}  
$$
and 
$$
\Zom:=\{\Phi=\{\phi_{k}\}_{k=1}^{\infty} \ | \ \phi_{k}\in
H_{0}^{\frac{1}{2}}(\Omega)\cap H^{1}(\Omega), \ \forall \ k\in {\mathbb N}, \ 
\|\Phi\|_{\Zom}<\infty \}  
$$
respectively. We also introduce the generalized homogenous Sobolev norms 
$$
\|\Phi\|_{\dYom}:=(\sum_{k=1}^{\infty}\lambda_{k}{\||p|^{\frac{1}{2}}
\phi_{k}\|_{L^{2}(\Omega)}^{2}})^{\frac{1}{2}} \ and \
\|\Phi\|_{\dZom}:=(\sum_{k=1}^{\infty}\lambda_{k}{\|\nabla \phi_{k}\|
_{L^{2}(\Omega)}^{2}})^{\frac{1}{2}}.
$$
Here, $|p|$ stands for the operator $\sqrt{-\Delta}$,  and has the meaning of the relativistic
kinetic energy of a particle with zero mass.
We note the following equivalence of norms.

\bigskip

\begin{lemma}\label{lm:2}
For $\Phi\in \Yom,$ the norms 
$\|\Phi\|_{\Yom}$ and $\|\Phi\|_{\dYom}$ are equivalent. 
If $\Phi\in \Zom,$ then
$\|\Phi\|_{\Zom}$ is equivalent to  $\|\Phi\|_{\dZom}$.
\end{lemma}

\bigskip

\prf Clearly
$$ 
\|\Phi\|_{\dYom}\leq 
(\sum_{k=1}^{\infty}\lambda_{k}\{\|\phi_{k}\|_{L^{2}(\Omega)}^{2} +
\||p|^{\frac{1}{2}}\phi_{k}\|_{L^{2}(\Omega)}^{2}\})^{\frac{1}{2}}=
(\sum_{k=1}^{\infty}\lambda_{k}{\|\phi_{k}\|
_{H^{\frac{1}{2}}(\Omega)}^{2}})^{\frac{1}{2}}=\|\Phi \|_{\Yom}.
$$
We will make use of the Poincar\'e inequality
\begin{equation}
\label{P}
\int_{\Omega}|\nabla \phi_{k}|^{2}dx\geq c_{p}\int_{\Omega}|\phi_{k}|^{2}dx
\end{equation}
with the constant $c_{p}>0$ dependent upon the domain $\Omega$ with Dirichlet
boundary conditions. Thus 
$$
\||p|^{\frac{1}{2}}\phi_{k}\|_{L^{2}(\Omega)}^{2}\geq \sqrt{c_{p}}
\|\phi_{k}\|_{L^{2}(\Omega)}^{2},
$$
which enables us to estimate
$$
\|\Phi \|_{\Yom}=(\sum_{k=1}^{\infty}\lambda_{k}\{\|\phi_{k}\|
_{L^{2}(\Omega)}^{2} +\||p|^{\frac{1}{2}}\phi_{k}\|_{L^{2}(\Omega)}^{2}\})
^{\frac{1}{2}}\leq
$$
$$ 
\leq \sqrt{1+{\frac{1}{\sqrt{c_{p}}}}}
(\sum_{k=1}^{\infty}\lambda_{k}{\||p|^{\frac{1}{2}}
\phi_{k}\|_{L^{2}(\Omega)}^{2}})^{\frac{1}{2}}=C\|\Phi\|_{\dYom}.
$$
Let us compare the remaining two norms. Clearly,
$$
\|\Phi \|_{\dZom}\leq (\sum_{k=1}^{\infty}\lambda_{k}\|\phi_{k}\|_
{H^{1}(\Omega)}^{2})^{\frac{1}{2}}=\|\Phi \|_{\Zom}.
$$
On the other hand, by means of the Poincar\'e inequality (\ref{P}),
$$
\|\Phi \|_{\Zom}=(\sum_{k=1}^{\infty}\lambda_{k}\{\|\phi_{k}\|
_{L^{2}(\Omega)}^{2} +\|\nabla\phi_{k}\|_{L^{2}(\Omega)}^{2}\})
^{\frac{1}{2}}\leq
$$
$$
\leq \sqrt{1+{\frac{1}{c_{p}}}}(\sum_{k=1}^{\infty}
\lambda_{k}{\|\nabla \phi_{k}\|_{L^{2}(\Omega)}^{2}})^{\frac{1}{2}}=
\|\Phi \|_{\dZom}.
$$
\endprf

\bigskip

Let $\Psi=\{\psi_{m}\}_{m=1}^{\infty}$ be a wave function and the relativistic
kinetic energy operator acts on it $T_{m}\Psi=(\sqrt{-\Delta+m^{2}}-m)\psi$  
componentwise. We have the following two lemmas.

\bigskip

\begin{lemma}\label{lm:3}  
The domain of the kinetic energy operator is given by
$D(T_{m})=\Zom\subseteq \Xom$.
\end{lemma}

\bigskip

\prf Let $\Psi \in \Zom$. Then 
$$
\sum_{m=1}^{\infty}\lambda_{m}\|\psi_{m}\|_{H^{1}(\Omega)}^{2}=
\sum_{m=1}^{\infty}\lambda_{m}\{\|\psi_{m}\|_{L^{2}(\Omega)}^{2}+
\|\nabla \psi_{m}\|_{L^{2}(\Omega)}^{2}\}\geq
\sum_{m=1}^{\infty}\lambda_{m}\|\psi_{m}\|_{L^{2}(\Omega)}^{2},
$$
and also, $\|\Psi\|_{\Xom}<\infty$. We estimate
$$
\|T_{m}\psi_{k}\|_{L^{2}(\Omega)}^{2}=((-\Delta+m^{2})\psi_{k}, \psi_{k})
_{L^{2}(\Omega)}+m^{2}\|\psi_{k}\|_{L^{2}(\Omega)}^{2}-
2m(\sqrt{-\Delta+m^{2}}\psi_{k}, \psi_{k})_{L^{2}(\Omega)}\leq 
$$
$$
\leq \|\nabla \psi_{k}\|_{L^{2}(\Omega)}^{2}+2m^{2}\|\psi_{k}\|
_{L^{2}(\Omega)}^{2}\leq c(m)\|\psi_{k}\|_{H^{1}(\Omega)}^{2},  
$$
where $c(m)$ is a mass dependent constant. Hence 
$$
\|T_{m}\Psi\|_{\Xom}^{2}=\sum_{k=1}^{\infty}\lambda_{k}
\|T_{m}\psi_{k}\|_{L^{2}(\Omega)}^{2}\leq c(m)\sum_{k=1}^{\infty}\lambda_{k}
\|\psi_{k}\|_{H^{1}(\Omega)}^{2}<\infty.
$$
\endprf

\bigskip

\begin{lemma}\label{lm:4} 
The operator $T_{m}$ generates the group $
e^{-iT_{m}t}, \ t\in {\mathbb R}$, of unitary operators on $\Xom$.
\end{lemma}

\bigskip

\prf For $\alpha, \beta\in \Xom$ we compute the inner product
$$
(e^{-iT_{m}t}\alpha,e^{-iT_{m}t}\beta)_{\Xom}=\sum_{k=1}^{\infty}
\lambda_{k}(e^{-iT_{m}t}\alpha_{k},e^{-iT_{m}t}\beta_{k})_{L^{2}(\Omega)}=
\sum_{k=1}^{\infty}\lambda_{k}(\alpha_{k},\beta_{k})_{L^{2}(\Omega)}=
(\alpha,\beta)_{\Xom}.  
$$
\endprf

\bigskip

We rewrite the Schr\"odinger-Poisson system for $x\in \Omega$ into the form
\begin{equation}
\label{Sch}
\Psi_{t}=-iT_{m}\Psi+F[\Psi(x,t)], \ where \ F[\Psi]:=i^{-1}V[\Psi]\Psi,
\end{equation}
$$
-\Delta V[\Psi]=n[\Psi],  \ where \ V|_{\partial \Omega}=0,
$$
$$
n[\Psi]=\sum_{k=1}^{\infty}\lambda_{k}|\psi_{k}|^{2}
$$
and prove the following auxiliary result.

\bigskip

\begin{lemma}\label{lm:5} 
The map defined in (\ref{Sch}) $F: \Zom\to \Zom$ 
is locally Lipschitz continuous. 
\end{lemma}

\bigskip

\prf Let $\Psi, \ \Phi\in \Zom$ with 
$\Psi=\{\psi_{k}\}_{k=1}^{\infty}, \ \Phi=\{\phi_{k}\}_{k=1}^{\infty}$ and
$t\in [0,T]$. Then,
$$
\|F[\Psi]-F[\Phi]\|_{\Zom}=\|i^{-1}V[\Psi]\Psi-i^{-1}V[\Phi]\Phi
\|_{\Zom}=\|V[\Psi](\Psi-\Phi)+(V[\Psi]-V[\Phi])\Phi\|_
{\Zom}.
$$
This can be easily estimated above by means of Lemma \ref{lm:2} by
$$
C\|V[\Psi](\Psi-\Phi)\|_{\dZom}+
C\|(V[\Psi]-V[\Phi])\Phi \|_{\dZom},
$$
which equals 
\begin{equation}
\label{nabla}
C(\sum_{k=1}^{\infty}\lambda_{k}\|\nabla(V[\Psi](\psi_{k}-\phi_{k}))\|_
{L^{2}(\Omega)}^{2})^{\frac{1}{2}}+
C(\sum_{k=1}^{\infty}\lambda_{k}\|\nabla((V[\Psi]-V[\Phi])\phi_{k})\|_
{L^{2}(\Omega)}^{2})^{\frac{1}{2}}.
\end{equation}
Here, $C$  denotes a finite, positive, universal constant.
Clearly, we have
$$
\|\nabla(V[\Psi](\psi_{k}-\phi_{k}))\|_{L^{2}(\Omega)}^{2}\leq 
2\|(\nabla V[\Psi])(\psi_{k}-\phi_{k})\|_{L^{2}(\Omega)}^{2}+
2\|V[\Psi]\nabla(\psi_{k}-\phi_{k})\|_{L^{2}(\Omega)}^{2}.
$$
By means of the Schwarz inequality this can be bounded above by
$$
C\|\nabla V[\Psi]\|_{L^{4}(\Omega)}^{2}\|\psi_{k}-\phi_{k}\|_{L^{6}(\Omega)}^{2}+
2\|V[\Psi]\|_{L^{\infty}(\Omega)}^{2}\|\nabla(\psi_{k}-\phi_{k})\|_{L^{2}(\Omega)}^{2}.
$$
By applying the Sobolev embedding theorems to
these expressions, we arrive at
$$
C\|\Delta V[\Psi]\|_{L^{2}(\Omega)}^{2}\|\nabla(\psi_{k}-\phi_{k})\|
_{L^{2}(\Omega)}^{2}\leq C\|V[\Psi]\|_{H^{2}(\Omega)}^{2}\|\nabla(\psi_{k}-\phi_{k})\|
_{L^{2}(\Omega)}^{2}.
$$
To estimate the remaining term in (\ref{nabla}), we use
$$
\|\nabla ((V[\Psi]-V[\Phi])\phi_{k})\|_{L^{2}(\Omega)}^{2}\leq 
2\|\nabla (V[\Psi]-V[\Phi])\phi_{k}\|_{L^{2}(\Omega)}^{2}+
2\|(V[\Psi]-V[\Phi])\nabla \phi_{k}\|_{L^{2}(\Omega)}^{2}.
$$
The Schwarz inequality yields 
$$
2\|\nabla (V[\Psi]-V[\Phi])\|_{L^{4}(\Omega)}^{2}\|\phi_{k}\|_{L^{4}(\Omega)}^{2}+
2\|(V[\Psi]-V[\Phi])\|_{L^{\infty}(\Omega)}^{2}\|\nabla\phi_{k}\|_{L^{2}(\Omega)}^{2}.
$$
Applying the Sobolev embedding theorem along with the H\"older inequality
to these expressions, we find
$$
C\|\Delta (V[\Psi]-V[\Phi])\|_{L^{2}(\Omega)}^{2}\|\phi_{k}\|_{L^{6}(\Omega)}^{2}+
C\|\Delta (V[\Psi]-V[\Phi])\|_{L^{2}(\Omega)}^{2}\|\nabla \phi_{k}\|_
{L^{2}(\Omega)}^{2}.
$$
From the Sobolev inequality used in the first of the two terms above we 
deduce the upper bound
$$
C\|V[\Psi]-V[\Phi]\|_{H^{2}(\Omega)}^{2}\|\nabla\phi_{k}\|_{L^{2}(\Omega)}^{2}.
$$
Therefore, for the norm of the difference $\|F[\psi]-F[\Phi]\|_{\Zom}$
we have the estimate from above as
$$
C\|V[\Psi]\|_{H^{2}(\Omega)}(\sum_{k=1}^{\infty}
\lambda_{k}\|\nabla(\psi_{k}-\phi_{k})\|_{L^{2}(\Omega)}^{2})^{\frac{1}{2}}+
C\|V[\Psi]-V[\Phi]\|_{H^{2}(\Omega)}
(\sum_{k=1}^{\infty}\lambda_{k}\|\nabla \phi_{k}\|_{L^{2}(\Omega)}^{2})^{\frac{1}{2}},
$$
which obviously equals to
$$
C\|V[\Psi]\|_{H^{2}(\Omega)}\|\Psi-\Phi\|_{\dZom}+C\|V[\Psi]-V[\Phi]\|
_{H^{2}(\Omega)}\|\Phi\|_{\dZom}.
$$
Let us apply the Poincar\'e and the Schwarz inequalities to estimate the Sobolev
norm of the potential function as
$$
\|V[\Psi]\|_{H^{2}(\Omega)}\leq C\|\Delta V\|_{L^{2}(\Omega)}=
C\|n[\Psi]\|_{L^{2}(\Omega)}.
$$
Hence, our goal is to estimate the appropriate norm of the particle 
concentration. From the Schwarz inequality,
$$
\|n[\Psi]\|_{L^{2}(\Omega)}^{2}=\sum_{k,l=1}^{\infty}\lambda_{k}\lambda_{l}
(|\psi_{k}|^{2},|\psi_{l}|^{2})_{L^{2}(\Omega)}\leq
(\sum_{k=1}^{\infty}\lambda_{k}\|\psi_{k}\|_{L^{4}(\Omega)}^{2})^{2}.
$$
and using the H\"older inequality along with the Sobolev inequality,
$$
\|n[\Psi]\|_{L^{2}(\Omega)}\leq C \sum_{k=1}^{\infty}\lambda_{k}
\|\psi_{k}\|_{L^{6}(\Omega)}^{2}\leq C \sum_{k=1}^{\infty}\lambda_{k}
\|\nabla \psi_{k}\|_{L^{2}(\Omega)}^{2}.
$$
Hence, we arrive at the estimates for the particle concentration and the norms on the
potential function,
$$
\|n[\Psi]\|_{L^{2}(\Omega)}\leq C\|\Psi\|_{\dZom}^{2}, \quad
\|V[\Psi]\|_{H^{2}(\Omega)}\leq C\|\Psi\|_{\dZom}^{2}
$$
with $\|\,\cdot\,\|_{\dZom}$ and $\|\,\cdot\,\|_{\Zom}$ equivalent via Lemma \ref{lm:2}.
Evidently,
$$
W:=V[\Psi]-V[\Phi]
$$ 
satisfies the Poisson equation,
$$
-\Delta W=n[\Psi]-n[\Phi], \quad W|_{\partial \Omega}=0,
$$ 
and  Dirichlet boundary conditions.
Applying the Poincar\'e inequality along with the Schwarz inequality,
we arrive at
$$
\|W\|_{H^{2}(\Omega)}^{2}\leq C \|\Delta W\|_{L^{2}(\Omega)}^{2},
$$ 
such that 
$$
\|W\|_{H^{2}(\Omega)}\leq C\|n[\Psi]-n[\Phi]\|_{L^{2}(\Omega)}.
$$
We will use the trivial inequality
$$
|n[\Psi]-n[\Phi]|\leq \sum_{k=1}^{\infty}\lambda_{k}(|\psi_{k}|+|\phi_{k}|)
|\psi_{k}-\phi_{k}|.
$$ 
The Schwarz inequality applied twice yields
$$
\|n[\Psi]-n[\Phi]\|_{L^{2}(\Omega)}^{2}\leq \Bigg(\sum_{k=1}^{\infty}\lambda_{k}
\sqrt{\int_{\Omega}(|\psi_{k}|+|\phi_{k}|)^{2}|\psi_{k}-\phi_{k}|^{2}dx}\Bigg)
^{2}\leq 
$$
$$
\leq (\sum_{k=1}^{\infty}\lambda_{k}\||\psi_{k}|+|\phi_{k}|\|_{L^{4}(\Omega)}
\|\psi_{k}-\phi_{k}\|_{L^{4}(\Omega)})^{2}\leq 
(\sum_{k=1}^{\infty}\lambda_{k}(\|\psi_{k}\|_{L^{4}(\Omega)} +\|\phi_{k}|\|
_{L^{4}(\Omega)})\|\psi_{k}-\phi_{k}\|_{L^{4}(\Omega)})^{2},
$$
and using it again gives  
$$
\sum_{k=1}^{\infty}\lambda_{k}(\|\psi_{k}\|_{L^{4}(\Omega)} +\|\phi_{k}|\|
_{L^{4}(\Omega)})^{2}\sum_{s=1}^{\infty}\lambda_{s}\|\psi_{s}-\phi_{s}\|_
{L^{4}(\Omega)}^{2}.
$$
Applying the H\"older and Sobolev inequalities, we arrive at
$$
C\sum_{k=1}^{\infty}\lambda_{k}(\|\nabla \psi_{k}\|_{L^{2}(\Omega)}^{2}+
\|\nabla \phi_{k}\|_{L^{2}(\Omega)}^{2})\sum_{s=1}^{\infty}\lambda_{s}
\|\nabla \psi_{s}-\nabla \phi_{s}\|_{L^{2}(\Omega)}^{2}.
$$
This quantity can be easily estimated above by
$$
C\Bigg(\sum_{k=1}^{\infty}\lambda_{k}\|\psi_{k}\|_{H^{1}(\Omega)}^{2}+
\sum_{l=1}^{\infty}\lambda_{l}\|\phi_{l}\|_{H^{1}(\Omega)}^{2}\Bigg)\sum_{s=1}^{\infty}
\lambda_{s}\|\psi_{s}-\phi_{s}\|_{H^{1}(\Omega)}^{2},
$$ 
which clearly equals to
$$
C(\|\Psi\|_{\Zom}^{2}+\|\Phi\|_{\Zom}^{2})\|\Psi-\Phi\|_{\Zom}^{2}.
$$
Therefore,
$$
\|n[\Psi]-n[\Phi]\|_{L^{2}(\Omega)}\leq C(\|\Psi\|_{\Zom}+\|\Phi\|_{\Zom})
\|\Psi-\Phi\|_{\Zom}
$$
and
$$
\|V[\Psi]-V[\Phi]\|_{H^{2}(\Omega)}\leq C(\|\Psi\|_{\Zom}+\|\Phi\|_{\Zom})
\|\Psi-\Phi\|_{\Zom}.
$$
Collecting the estimates above, we arrive at
$$
\|F[\Psi]-F[\Phi]\|_{\Zom}\leq C(\|\Psi\|_{\Zom}^{2}+\|\Phi\|_
{\Zom}^{2})\|\Psi-\Phi\|_{\Zom},
$$
which completes the proof of the lemma. \endprf

\bigskip

From standard arguments   (see for instance  Theorem 1.7 of \cite{P83}) thus follows
that the above Schr\"odinger-Poisson system admits a unique mild solution 
$\Psi$ in $\Zom$ on a time interval $[0, T)$, for some $T>0$, satisfying the integral equation
\eqn\label{eq:mildsol}
	\Psi(t)=e^{-iT_m t} \Psi(0) + e^{-iT_mt} \int_0^t e^{iT_ms}F[\Psi(s)]ds
\eeqn
in $\Zom$. Moreover,
$$
\hbox{lim}_{t\nearrow T}\|\Psi(t)\|_{\Zom}=\infty
$$
if $T$ is finite. 
We also note that $\Psi$ is a unique strong solution in $\Xom$.
We shall next prove that this solution is in fact global in time.
First we prove the following lemma.

\bigskip 

\begin{lemma}\label{lm:6} 
Suppose for the unique mild solution \eqref{eq:mildsol} of the Schr\"odinger-Poisson system 
(\ref{sch})-(\ref{n}) that  
$\{\psi_{k}(x, 0)\}_{k=1}^{\infty}$ at $t=0$
forms a complete orthonormal system in $L^{2}(\Omega)$. Then, for any 
$t\in [0, T),$ the set
$\{\psi_{k}(x, t)\}_{k=1}^{\infty}$ remains  a complete orthonormal system 
in $L^{2}(\Omega)$. Moreover, the $\Xom$-norm is preserved,
$\|\Psi(x,t)\|_{\Xom}=\|\Psi(x,0)\|_{\Xom}, \ t\in [0,T)$.
\end{lemma}

\bigskip

\prf 
Given the solution $\Psi(t)$ of the Schr\"odinger-Poisson system on $[0,T)$, we obtain the 
time-dependent one-particle Hamiltonian
$$
	H_{V_\Psi}(t) = T_m + V_\Psi(t,x)
$$
where the potential $V_\Psi$ solves 
$-\Delta V_\Psi(t,x)=n[\Psi(t)]$ with Dirichlet boundary conditions, see \eqref{eq:HV-def-1}. 
Accordingly, the components of $\Psi(t)$ solve the {\em linear}, {\em non-autonomous} Schr\"odinger equation $i\partial_t\psi_k(t,x)=H_{V_\Psi}(t)\psi_k(t,x)$, for $k\in\N$,
on the time interval  $[0,T)$. We thus have, for $t\in[0,T)$,
\begin{equation}
\label{ev} 
\psi_{k}(x, t)=(e^{-i\int_{0}^{t}H_{V_\Psi}(\tau)d\tau}\psi_{k})(x, 0), \ k\in {\mathbb N},
\end{equation}
and therefore
$$
(\psi_{k}(x, t),\psi_{l}(x, t))_{L^{2}(\Omega)}=
(e^{-i\int_{0}^{t}H_{V_\Psi}(\tau)d\tau}\psi_{k}(x, 0),e^{-i\int_{0}^{t}H_{V_\Psi}(\tau)d\tau}\psi_{l}
(x, 0))_{L^{2}(\Omega)}=
$$
$$
=(\psi_{k}(x, 0),\psi_{l}(x, 0))_{L^{2}(\Omega)}=\delta_{k,l}, \quad k,l\in 
{\mathbb N},
$$
where $\delta_{k,l}$ stands for the Kronecker symbol. Obviously, for 
$k\in {\mathbb N}$,
$$
\|\psi_{k}(x, t)\|_{L^{2}(\Omega)}^{2}=\|\psi_{k}(x, 0)\|_{L^{2}(\Omega)}^{2},
$$
such that for $t\in [0,T)$, the $\Xom$-norm is conserved,
$$
\|\Psi(x,t)\|_{\Xom}=
(\sum_{k=1}^{\infty}\lambda_{k}\|\psi_{k}(x, t)\|_{L^{2}(\Omega)}^{2})^{\frac{1}{2}}=
(\sum_{k=1}^{\infty}\lambda_{k}\|\psi_{k}(x, 0)\|_{L^{2}(\Omega)}^{2})^{\frac{1}{2}}
=\|\Psi(x,0)\|_{\Xom}.
$$
Let us consider an 
arbitrary function $f(x)\in L^{2}(\Omega)$. Clearly, we have the expansion
$$
f(x)=\sum_{k=1}^{\infty}(f(y),\psi_{k}(y, 0))_{L^{2}(\Omega)}\psi_{k}(x, 0)
$$
and similarly
$$
e^{i\int_{0}^{t}H_{V_\Psi}(\tau)d\tau}f(x)=\sum_{k=1}^{\infty}(e^{i\int_{0}^{t}H_{V_\Psi}(\tau)d\tau}
f(y),\psi_{k}(y, 0))_{L^{2}(\Omega)}\psi_{k}(x, 0).
$$
Thus, by means of (\ref{ev}) we arrive at the expansion
$$
f(x)=\sum_{k=1}^{\infty}(f(y),\psi_{k}(y, t))_{L^{2}(\Omega)}\psi_{k}(x, t)
$$
for $t\in [0,T)$. \endprf

\bigskip

Furthermore, we have conservation of energy for solutions to the Schr\"odinger-Poisson system
in the following sense.

\bigskip

\begin{lemma}\label{lm:7} 
For the unique mild solution \eqref{eq:mildsol} of the Schr\"odinger-Poisson
system (\ref{sch})-(\ref{n}) and for any value of time $t\in [0,T)$ we have 
the identity
$$
\|\Psi(x,t)\|_{\dYom}^{2}+\frac{1}{2}\|\nabla V[\Psi(x,t)]\|_{L^{2}(\Omega)}^{2}=
\|\Psi(x,0)\|_{\dYom}^{2}+\frac{1}{2}\|\nabla V[\Psi(x,0)]\|_{L^{2}(\Omega)}^{2}.
$$
\end{lemma}

\prf Complex conjugation of the  
Schr\"odinger-Poisson system  (\ref{sch}) yields
\begin{equation}
\label{conj}
-i\frac{\partial \bar{\psi}_{k}}{\partial t}=T_{m}\bar{\psi}_{k}+V[\psi]
\bar{\psi}_{k}, \quad k\in {\mathbb N}.
\end{equation}
Adding the $k$-th equation of the original system (\ref{sch}) multiplied by 
$\displaystyle{\frac{\partial \bar{\psi}_{k}}{\partial t}}$, 
and the $k$-th equation
in (\ref{conj}) multiplied by $\displaystyle{\frac{\partial \psi_{k}}{\partial t}}$,  we  obtain
$$
\frac{\partial}{\partial t}\|T_{m}^{\frac{1}{2}}\psi_{k}\|_{L^{2}(\Omega)}^{2}+
\int_{\Omega}V[\psi]\frac{\partial}{\partial t}|\psi_{k}|^{2}dx=0, \quad k\in 
{\mathbb N}.
$$
Thus, multiplying by $\lambda_{k}$, and summing over $k$, we find  
\begin{equation}
\label{enc}
\frac{\partial}{\partial t}\|\Psi(x,t)\|_{\dYom}^{2}+\int_{\Omega}V[\Psi(x,t)]
\frac{\partial}{\partial t}n[\Psi(x,t)]dx=0.
\end{equation}
One can easily verify the identity
$$
\frac{\partial}{\partial t}\|\nabla V[\Psi(x,t)]\|_{L^{2}({\Omega})}^{2}=2
\int_{\Omega}V[\Psi(x,t)]\frac{\partial}{\partial t}n[\Psi(x,t)]dx,
$$
which we substitute in (\ref{enc}) to complete the proof of the lemma.
\endprf

\bigskip

With the auxiliary statements proven above at our disposal, we may now
prove our main result, Theorem \ref{thm:1}.

\bigskip

\prf[Proof of Theorem \ref{thm:1}]
The proof follows from the blow-up alternative and conservation laws. It follows from Lemma \ref{lm:7} that $\|\Psi(t)\|_{\dYom}$ is bounded from above uniformly in time,
$$\|\Psi(t)\|^2_{\dYom} \le \|\Psi(t)\|_{\dYom}^{2}+\frac{1}{2}\|\nabla V[\Psi(t)]\|_{L^{2}(\Omega)}^{2}=
\|\Psi(0)\|_{\dYom}^{2}+\frac{1}{2}\|\nabla V[\Psi(0)]\|_{L^{2}(\Omega)}^{2}.$$ 
We need to bound $\|\Psi(t)\|_{\dZom}.$
We recall  the mild solution of the Schr\"odinger-Poisson system (\ref{sch})-(\ref{n}), given by
\eqn\label{eq:mildsol-1}
	\Psi(t)=e^{-iT_m t} \Psi(0) + e^{-iT_mt} \int_0^t e^{iT_ms}F[\Psi(s)]ds,
\eeqn
which implies
$$\|\Psi(t)\|_{\Zom} \le \|\Psi (0)\|_{\Zom} + \int_0^t \|F[\Psi(s)]\|_{\Zom}.$$
From Lemma \ref{lm:2}, we have 
\begin{align*}
\|F[\Psi]\|_{\Zom} & = \|V[\Psi]\Psi\|_{\Zom} \le C \|V[\Psi]\Psi\|_{\dZom} \\
&\le C \left( \sum_{k=1}^\infty \lambda_k \|\nabla (V[\Psi]\psi_k)\|_{L^2(\Omega)}^2\right)^{1/2}.
\end{align*}
Now,
\begin{align*}
\|\nabla (V[\psi]\psi)\|_{L^2(\Omega)}^2&\le \|\nabla V[\Psi]\psi_k\|_{L^2(\Omega)}^2 + \|V[\Psi]\nabla\psi_k\|_{L^2(\Omega)}^2\\
&\le \|\nabla V[\Psi]\|_{L^6(\Omega)}^2 \|\psi_k\|_{L^3(\Omega)}^2 + \|V[\Psi]\|_{L^\infty(\Omega)}^2\|\nabla \psi_k\|_{L^2(\Omega)}^2 \\
&\le  \|\nabla V[\Psi]\|_{L^6(\Omega)}^2 \|\psi_k\|_{H^{1/2}(\Omega)}^2 + \|V[\Psi]\|_{L^\infty(\Omega)}^2\|\psi_k\|_{H^1(\Omega)}^2, 
\end{align*}
where we have used H\"older's inequality in the second line and the Sobolev inequality 
$$\|f\|_{L^{\frac{6}{3-2p}}(\Omega)}\le C \|f\|_{H^p(\Omega)}$$
in the last line.
To evaluate $\|\nabla V[\Psi]\|_{L^6(\Omega)},$ recall that $\Delta V[\Psi] = -n[\Psi].$
Applying H\"older's and Sobolev inequalities, we get
\begin{align*}
\|\nabla V[\Psi]\|_{L^6(\Omega)}^2 &\le C \|\nabla V[\Psi]\|_{H^1(\Omega)}^2\le C \|n[\Psi]\|_{L^2(\Omega)}^2 \\
&\le C\sum_{k,l=1}^\infty \lambda_k\lambda_l(|\psi_k|^2 , |\psi_l|^2)_{L^2(\Omega)} \le C \sum_{k,l=1}^\infty \lambda_k\lambda_l \|\psi_k\psi_l\|_{L^2(\Omega)}^2 \\
&\le C \sum_{k,l=1}^\infty \lambda_k\lambda_l \|\psi_k\|_{L^6(\Omega)}^2 \|\psi_l\|_{L^3(\Omega)}^2\le C (\sum_{k=1}^\infty \lambda_k \|\psi_k\|_{H^1(\Omega)}^2) (\sum_{l=1}^\infty \lambda_l \|\psi_l\|^2_{H^{1/2}(\Omega)})\\
&\le C \|\Psi\|_{\dZom}^2 \|\Psi\|_{\dYom}^2.
\end{align*}
We now estimate $\|V[\Psi]\|_{L^\infty(\Omega)}.$ The Sobolev inequality implies
$$\|V[\Psi]\|_{L^\infty(\Omega)}^2 \le C \| |p|^{-1/2} n[\Psi]\|_{L^2(\Omega)}^2.$$
We claim that $\||p|^{-1/2}n[\Psi]\|_{L^2(\Omega)}$ is controlled by $\|\Psi\|_{\dYom}.$
\begin{align*}
\| |p|^{-1/2} n[\Psi]\|_{L^2(\Omega)}^2 &= (n[\Psi], |p|^{-1}n[\Psi])_{L^2(\Omega)}\le \|n[\Psi]\|_{L^{3/2}(\Omega)} \||p|^{-1}n[\Psi]\|_{L^3(\Omega)} \\
&\le C \|\Psi\|_{L^3(\Omega)}^2 \||p|^{-1}n[\Psi]\|_{H^{1/2}(\Omega)} \le C \|\Psi\|_{H^{1/2}(\Omega)}^2 \||p|^{-1/2}n[\Psi]\|_{L^2(\Omega)},
\end{align*}
where we have used H\"older's inequality in the first line, and the Sobolev inequality in the second line. 
It follows that 
$$\||p|^{-1/2}n[\Psi]\|_{L^2(\Omega)} \le C \|\Psi\|_{\dYom}^2,$$
and hence
$$\|V[\Psi]\|_{L^\infty(\Omega)}^2 \le C \|\Psi\|_{\dYom}^4.$$
Combining the above estimates yields
$$\|F[\Psi]\|_{\dZom} \le C \|\Psi\|_{\dYom}^2 \|\Psi\|_{\dZom}.$$
This implies
$$\|\Psi(t)\|_{\dZom} \le \|\Psi(0)\|_{\dZom} + \int_0^t C_0 \|\Psi(s)\|_{\dZom},$$
where $C_0$ is a constant proportional to the initial energy $\|\Psi(0)\|_{\dYom}^{2}+\frac{1}{2}\|\nabla V[\Psi(0)]\|_{L^{2}(\Omega)}^{2}.$
By Gronwall's lemma, 
$$\|\Psi(t)\|_{\dZom} \le C_1e^{C_2 t}, \ \ t>0.$$
By the blow-up alternative, this implies that the Schr\"odinger-Poisson system is globally 
well-posed in $\Zom.$ 
\endprf

\bigskip

\noindent
{\bf Acknowledgements}

\bigskip

V.V. thanks J. Colliander, R. Jerrard, and I.M. Sigal for stimulating discussions.
In particular, the idea to pursue this problem originates from discussions and 
seminars  on related topics  initiated by I.M. Sigal that V.V. was part of, 
around 2007  at the University of Toronto.
The work of T.C. was supported by NSF grant DMS-1009448. 
W.A.S. acknowledges the support of NSERC discovery grant and USASK start-up fund.

\end{document}